\documentclass[
    ,final            
  ]
  {aipproc}

\layoutstyle{6x9}

\def\kpi{K^-\pi^+}
\def\kpipiz{K^-\pi^+\pi^0}
\def\kpipipi{K^-\pi^+\pi^+\pi^-}
\def\kpipi{K^-\pi^+\pi^+}
\def\kpipipiz{K^-\pi^+\pi^+\pi^0}
\def\kspi{K^0_S\pi^+}
\def\kspipiz{K^0_S\pi^+\pi^0}
\def\kspipipi{K^0_S\pi^+\pi^+\pi^-}
\def\kkpi{K^+K^-\pi^+}

\def\Dzkpi{D^0\to K^-\pi^+}

\def\Dpkpipi{D^+\to K^-\pi^+\pi^+}

\def\NDDbar{N_{D\bar D}}
\def\NDzDzbar{N_{D^0\bar D^0}}
\def\NDpDm{N_{D^+D^-}}

\begin{document}

\title{Measurements of Absolute Hadronic Branching Fractions of $D$ Mesons}


\author{Werner M. Sun, for the CLEO Collaboration}{
  address={Cornell University, Ithaca, New York 14853}
}

\begin{abstract}
Using $e^+e^-$ collisions recorded at the $\psi(3770)$
resonance with the CLEO-c detector at the Cornell Electron Storage Ring,
we determine absolute hadronic branching fractions of charged and neutral
$D$ mesons. Among measurements for both Cabibbo-favored and Cabibbo-suppressed
modes, we obtain reference branching fractions
${\cal B}(D^0\to K^-\pi^+)=(3.91\pm 0.08\pm 0.09)\%$ and
${\cal B}(D^+\to K^-\pi^+\pi^+)=(9.5\pm 0.2\pm 0.3)\%$, where the
uncertainties are statistical and systematic, respectively.
Using a determination of the integrated luminosity, we also extract the
$e^+e^-\to D\bar D$ cross sections. 
\end{abstract}

\maketitle


Absolute measurements of hadronic charm meson branching fractions play a
central role in the study of the weak interaction because they serve to
normalize many $D$ and $B$ meson branching fractions, from which CKM matrix
elements are determined.  At CLEO-c, we have measured several charge-averaged
branching fractions listed in Tables~\ref{tab-dataResults}
and~\ref{tab-dataResults2}.  Two of these modes, $\Dzkpi$ and $\Dpkpipi$, are
particularly important because essentially all other $D^0$ and $D^+$ branching
fractions have been determined from ratios to one of these branching fractions.
Our data sample was produced in $e^+e^-$ collisions on the $\psi(3770)$
resonance at the Cornell Electron Storage Ring and collected with the
CLEO-c detector.

For the results in Table~\ref{tab-dataResults}, based on 55.8 ${\rm pb}^{-1}$
of integrated luminosity, we employ a double tagging
technique pioneered by MARK III~\cite{markiii-1, markiii-2}, which
 obviates the need for knowledge of the
luminosity or the $e^+e^-\to D\bar D$ production cross section.
A single reconstructed $D$ or $\bar D$ (called
single tag or ST) tags the event as either $D^0\bar D^0$ or $D^+D^-$.
Double tag (DT) events have both the $D$ and $\bar D$ reconstructed.
The measured ST and DT yields are assumed to be
$N_i=\epsilon_i{\cal B}_iN_{D\bar D}$ and
$N_{ij} = \epsilon_{ij}{\cal B}_i{\cal B}_jN_{D\bar D}$,
$\epsilon_i$ and $\epsilon_{ij}$ are ST and DT efficiencies,
${\cal B}_i$ is the branching fraction for mode $i$ (assuming no
$D^0$-$\bar D^0$ mixing or $CP$ violation) and $N_{D\bar D}$ is the 
number of produced $D\bar D$ pairs.
Thus, we can extract the ${\cal B}_i$ and $\NDDbar$, simultaneously for $D^0$
and $D^+$, with a least-squares procedure described in Ref.~\cite{brfit}.
We identify $D$ candidates by their beam-constrained mass,
$M \equiv\sqrt{E_{\rm beam}^2 - {\mathbf p}_D^2}$, and by
$\Delta E\equiv E_D - E_{\rm beam}$.
The ${\cal B}_i$ and $\NDDbar$ statistical uncertainties
are dominated by those of the DT yields, which we find to be
$2484\pm 51$ for $D^0$ and $1650\pm 42$ for $D^+$.


\begin{table}
\begin{tabular}{lccc}
\hline
\tablehead{1}{c}{b}{$D$ Decay Mode} &
	\tablehead{1}{c}{b}{Fitted ${\cal B}$ (\%)} &
	\tablehead{1}{c}{b}{PDG ${\cal B}$ (\%)} &
	\tablehead{1}{c}{b}{$\Delta_{\rm FSR}$} \\
\hline
$\kpi$        & $3.91\pm 0.08\pm 0.09$        & $3.80\pm 0.09$ & ~$-2.0\%$ \\
$\kpipiz$     & $14.9\pm 0.3\pm 0.5$          & $13.0\pm 0.8$ & ~$-0.8\%$ \\
$\kpipipi$    & $8.3\pm 0.2\pm 0.3$           & $7.46\pm 0.31$ & ~$-1.7\%$ \\
\hline
$\kpipi$      & $9.5\pm 0.2\pm 0.3$           & $9.2\pm 0.6$ & ~$-2.2\%$ \\
$\kpipipiz$   & $6.0\pm 0.2\pm 0.2$           & $6.5\pm 1.1$ & ~$-0.6\%$ \\
$\kspi$       & $1.55\pm 0.05\pm 0.06$        & $1.41\pm 0.10$ & ~$-1.8\%$ \\
$\kspipiz$    & $7.2\pm 0.2\pm 0.4$           & $4.9\pm 1.5$ & ~$-0.8\%$ \\
$\kspipipi$   & $3.2\pm 0.1\pm 0.2$           & $3.6\pm 0.5$ & ~$-1.4\%$ \\
$\kkpi$       & $0.97\pm 0.04\pm 0.04$        & $0.89\pm 0.08$ & ~$-0.9\%$ \\
\hline
\tablehead{1}{c}{b}{$D\bar D$ Yield} &
	\tablehead{2}{c}{b}{Fitted Value} &
	\tablehead{1}{c}{b}{$\Delta_{\rm FSR}$} \\
\hline
$\NDzDzbar$ & \multicolumn{2}{c}{$(2.01\pm 0.04\pm 0.02)\times 10^5$} &
	~$-0.2\%$ \\
$\NDpDm$ & \multicolumn{2}{c}{$(1.56\pm 0.04\pm 0.01)\times 10^5$} &
	~$-0.2\%$ \\
\hline
\end{tabular}
\caption{Fitted branching fractions and $D\bar D$ pair yields, along with the
fractional FSR corrections and comparisons to the Particle Data
Group~\protect{\cite{PDG}} fit results.  Uncertainties are
statistical and systematic, respectively.}
\label{tab-dataResults}
\end{table}


The results of the data fit are shown in Table~\ref{tab-dataResults}.
The $\chi^2$ of the fit is
28.1 for 52 degrees of freedom, corresponding to a confidence level of 99.7\%.
All nine branching fractions have comparable precision to
the current PDG averages.  We do not explicitly reconstruct FSR
photons, but because FSR is simulated in the samples used to calculate
efficiencies, our branching fractions are inclusive of
photons radiated from the final state
particles.  If no FSR were included in the simulations, then all the branching
fractions would change by $\Delta_{\rm FSR}$ in Table~\ref{tab-dataResults}.

We obtain the $e^+e^-\to D\bar D$ cross sections by scaling
$\NDzDzbar$ and $\NDpDm$ by the luminosity,
${\cal L} = (55.8 \pm 0.6)$ ${\rm pb}^{-1}$.  Thus, at
$E_{\rm cm}=3773$ MeV, we find peak cross sections of
$\sigma( e^+e^-\to D^0\bar D^0 ) = (3.60\pm 0.07^{+0.07}_{-0.05}) \ {\rm nb}$,
$\sigma( e^+e^-\to D^+ D^- ) = (2.79\pm 0.07^{+0.10}_{-0.04}) \ {\rm nb}$,
$\sigma( e^+e^-\to D\bar D ) = (6.39\pm 0.10^{+0.17}_{-0.08}) \ {\rm nb}$, and
$\sigma( e^+e^-\to D^+ D^- ) / \sigma( e^+e^-\to D^0\bar D^0 ) = 0.776\pm 0.024^{+0.014}_{-0.006}$,
where the uncertainties are statistical and systematic, respectively.
The systematic uncertainties include uncertainties on $\NDzDzbar$, $\NDpDm$,
and ${\cal L}$, as well as the effect of $E_{\rm cm}$ variations with respect
to the peak.  Our measured
cross sections are in good agreement with BES~\cite{bessigma} and higher than
those of MARK III~\cite{markiii-2}.

For the Cabibbo-suppressed branching fractions in Table~\ref{tab-dataResults2},
based on 281 ${\rm pb}^{-1}$ of integrated luminosity, we measure
ST yields only and determine branching ratios with respect to the reference
modes $D^0\to K^-\pi^+$ and $D^+\to K^-\pi^+\pi^+$.  Backgrounds from
Cabibbo-favored decays with $K^0_S\to\pi^+\pi^-$ are suppressed with a veto
on the $\pi^+\pi^-$ invariant mass.  Six of the modes in
Table~\ref{tab-dataResults2} are observed for the first time, and we obtain
absolute branching fractions by combining the PDG average~\cite{PDG} and our
results in Table~\ref{tab-dataResults} for the reference modes.  For the three
$D\to\pi\pi$ modes, we also find the ratio of the $\Delta I=3/2$ to
$\Delta I=1/2$ isospin amplitudes to be
$A_2/A_0 = 0.420\pm 0.014 {\rm (stat.)}\pm 0.010 {\rm (syst.)}$ and the
relative strong phase to be $\delta_I = (86.4\pm 2.8\pm 3.3)^\circ$, which
indicates a substantial contribution from final state interactions.

\begin{table}
\begin{tabular}{lccc}
\hline
\tablehead{1}{c}{b}{$D$ Decay Mode} &
	\tablehead{1}{c}{b}{${\cal B}/{\cal R}_{0/+}$ (\%)} &
	\tablehead{1}{c}{b}{${\cal B}$ ($10^{-3}$)} &
	\tablehead{1}{c}{b}{PDG ${\cal B}$ ($10^{-3}$)} \\
\hline
$\pi^+\pi^-$ &
	$3.62\pm 0.10\pm 0.07\pm 0.04$ &
	$1.39\pm 0.04\pm 0.04\pm 0.03\pm 0.01$ &
	$1.38\pm 0.05$ \\
$\pi^0\pi^0$ &
	$2.05\pm 0.13\pm 0.16\pm 0.02$ &
	$0.79\pm 0.05\pm 0.06\pm 0.01\pm 0.01$ &
	$0.84\pm 0.22$ \\
$\pi^+\pi^-\pi^0$ &
	$34.4\pm 0.5\pm 1.2\pm 0.3$ &
	$13.2\pm 0.2\pm 0.5\pm 0.2\pm 0.1$ &
	$11\pm 4$ \\
$\pi^+\pi^+\pi^-\pi^-$ &
	$19.1\pm 0.4\pm 0.6\pm 0.2$ &
	$7.3\pm 0.1\pm 0.3\pm 0.1\pm 0.1$ &
	$7.3\pm 0.5$ \\
$\pi^+\pi^-\pi^0\pi^0$ &
	$25.8\pm 1.5\pm 1.8\pm 0.3$ &
	$9.9\pm 0.6\pm 0.7\pm 0.2\pm 0.1$ \\
$\pi^+\pi^+\pi^-\pi^-\pi^0$ &
	$10.7\pm 1.2\pm 0.5\pm 0.1$ &
	$4.1\pm 0.5\pm 0.2\pm 0.1\pm 0.0$ \\
$\omega\pi^+\pi^-$ &
	$4.1\pm 1.2\pm 0.4\pm 0.0$ &
	$1.7\pm 0.5\pm 0.2\pm 0.0\pm 0.0$ \\
$\eta\pi^0$ &
	$1.47\pm 0.34\pm 0.11\pm 0.01$ &
	$0.62\pm 0.14\pm 0.05\pm 0.01\pm 0.01$ \\
$\pi^0\pi^0\pi^0$ &
	--- &
	$< 0.35$ (90\% C.L.) \\
$\omega\pi^0$ &
	--- &
	$< 0.26$ (90\% C.L.) \\
$\eta\pi^+\pi^-$ &
	--- &
	$< 1.9$ (90\% C.L.) \\
\hline
$\pi^+\pi^0$ &
	$1.33\pm 0.07\pm 0.06$ &
	$1.25\pm 0.06\pm 0.07\pm 0.04$ &
	$1.33\pm 0.22$ \\
$\pi^+\pi^+\pi^-$ &
	$3.52\pm 0.11\pm 0.12$ &
	$3.35\pm 0.10\pm 0.16\pm 0.12$ &
	$3.1\pm 0.4$ \\
$\pi^+\pi^0\pi^0$ &
	$5.0\pm 0.3\pm 0.3$ &
	$4.8\pm 0.3\pm 0.3\pm 0.2$ \\
$\pi^+\pi^+\pi^-\pi^0$ &
	$12.4\pm 0.5\pm 0.6$ &
	$11.6\pm 0.4\pm 0.6\pm 0.4$ \\
$\pi^+\pi^+\pi^+\pi^-\pi^-$ &
	$1.73\pm 0.20\pm 0.17$ &
	$1.60\pm 0.18\pm 0.16\pm 0.06$ &
	$1.73\pm 0.23$ \\
$\eta\pi^+$ &
	$3.81\pm 0.26\pm 0.21$ &
	$3.61\pm 0.25\pm 0.23\pm 0.12$ &
	$3.0\pm 0.6$ \\
$\omega\pi^+$ &
	--- &
	$< 0.34$ (90\% C.L.) \\
\hline
\end{tabular}
\caption{Ratios of branching fractions to the reference
branching fractions ${\cal R}_0\equiv {\cal B}(\Dzkpi)$ and
${\cal R}_+\equiv {\cal B}(\Dpkpipi)$, along with comparisions to the Particle
Data Group~\protect{\cite{PDG}} fit results.
Uncertainties arise from statistics, experimental systematic effects, 
${\cal R}_{0/+}$, and quantum correlations ($D^0$ modes
only)~\cite{Asner:2005wf}.  For the relative branching fractions, the
${\cal R}_{0/+}$ uncertainty is omitted.}
\label{tab-dataResults2}
\end{table}

Using the 281 ${\rm pb}^{-1}$ sample, we also search for an asymmetry between
${\cal B}(D^+\to K^0_S\pi^+)$ and ${\cal B}(D^+\to K^0_L\pi^+)$, which
can arise from interference among competing amplitudes~\cite{Bigi:1994aw}.
We reconstruct the neutral kaon inclusively by fully-reconstructing the $D^-$,
finding the $\pi^+$ daughter of the $D^+$, and computing the missing mass of
the event, which peaks at the neutral kaon mass for both $K^0_S\pi^+$ and
$K^0_L\pi^+$ signal decays.  The dominant background comes from
$D^+\to\eta\pi^+$, which partially overlaps with $K^0\pi^+$ in missing mass.
We find a branching fraction asymmetry of
$[{\cal B}(K^0_L\pi^+)-{\cal B}(K^0_S\pi^+)]/[{\cal B}(K^0_L\pi^+)+{\cal B}(K^0_S\pi^+)] = -0.01\pm 0.04\pm 0.07$, which is consistent with the prediction of
${\cal O}(10\%)$~\cite{Bigi:1994aw}.


\begin{theacknowledgments}
We gratefully acknowledge the effort of the CESR staff in providing
us with excellent luminosity and running conditions.  This work was
supported by the National Science Foundation and the U.S. Department
of Energy.
\end{theacknowledgments}






\end{document}